\documentclass[osajnl,twocolumn,showpacs,superscriptaddress,10pt]{revtex4-1} 
\usepackage{epsf,epsfig}
\usepackage[psamsfonts]{amssymb}

\usepackage{amsmath}
\usepackage{bm}
\usepackage{graphicx}
\usepackage{graphicx,epstopdf}
\usepackage{caption}
\usepackage{subcaption}

\begin{document}

\title{Influence of spherical anisotropy on optical mass sensing in a molecular-plasmonic optomechanical system }

\author{Elnaz Aleebrahim}
\affiliation{Department of Physics, University of Isfahan, Hezar Jerib St., Isfahan 81764-73441, Iran.}

\author{Malek Bagheri Harouni}\email{Corresponding author: m-bagheri@phys.ui.ac.ir}
\affiliation{Department of Physics, University of Isfahan, Hezar Jerib St., Isfahan 81764-73441, Iran.}
\affiliation{Quantum optics group, Department of Physics, University of
Isfahan, Hezar Jerib St., Isfahan 81764-73441, Iran.}

\author{Ehsan Amooghorban}\email{Corresponding author: Ehsan.amooghorban@sku.ac.ir}
\affiliation{ Department of Physics, Faculty of Basic Sciences, Shahrekord University, P.O. Box 115, Shahrekord 88186-34141, Iran.}
\affiliation{ Photonic Research Group, Shahrekord University, Shahrekord 88186-34141, Iran.}

\begin{abstract}
We use an all-optical pump-probe method to develop a mass sensing mechanism in a molecular plasmonic system at room temperature. The system consists of a double-clamped graphene nanoribbon that parametrically interacts with two types of isotropic and anisotropic spherical plasmonic cavities in the presence of a strong pump field and a weak probe pulse. Based on the mode-selective quantization scheme and analogy with the canonical model of the cavity optomechanics, we formulate the Hamiltonian of the system in terms of the electromagnetic Green's tensor.
In this manner,  we derive an explicit form of size-dependent optomechanical coupling function and plasmonic damping rate, which include the modal, geometrical, and material features of the plasmonic structure. Engineering material features of the plasmonic nanostructure, we find that the intensity of the probe field transmission spectrum for radially anisotropic spherical nanocavity enhances significantly compared to the silver sphere nanocavity due to the mode volume reduction. This scheme can provide to achieve the minimum measurable mass $\Delta m \approx 10^{- 24} kg$ at room temperature.
\end{abstract}

\maketitle

\section{Introduction}
One of the most promising methods for tailoring light-matter interaction is to employ metallic structures in nanoscale size as plasmonic nanocavities~\cite{Tame2013,Neuman2018,Hugall2018,Stockman2018,Chikkaraddy2016,Xiong2018}.
Localized plasmon resonances (LPR), which result from the confinement of light waves within the metallic subwavelength structures, are considered as the fundamental concept of a new growing field known as plasmonic cavity quantum electrodynamic (PCQED)~\cite{Zhang2012,Balytis2018,Ginzburg2016}.
These Localized plasmons can be effectively controlled by adjusting cavity composition, size, shape, and the material parameters of the surrounding medium. The sub-diffraction-limit focusing of electromagnetic fields in the near-field of the LPRs is responsible for the enormous enhancement of the Raman spectrum in a commonly sensing technique known as the surface-enhanced Raman scattering(SERS)~\cite{Ding2017}.
So far, the dynamical behavior of SERS phenomena, which represents the interaction between molecular species and metallic nanoparticles, was formulated theoretically in the field of molecular plasmonic in both classical and quantum regimes~\cite{Tame2013,Hugall2018,Chikkaraddy2016,Fleischmann1974}.
Many attempts have been devoted to investigate the influence of material and geometrical parameters such as anisotropy on the SERS enhancement factor by finding the scalar potential distribution of the nanostructure and extracting its polarizability in the quasi-static approximation (QSA)~\cite{Yin2011}.

Recently a theoretical approach is introduced to explain the dynamical nature of plasmon-phonon interaction by mapping the molecular plasmonic system onto the canonical model of cavity optomechanics~\cite{Roelli2016}. Based on this model, it is shown that the polarizability of the molecule and its interaction with a plasmonic nanocavity are explicitly dependent on the internal vibrational states of the adjacent molecule~\cite{Schmidt2017,Aspelmeyer2014}.
This theoretical approach was applied to the SERS phenomenon and successfully described fundamental classical properties such as the dependence of the Raman signal on the intensity and frequency of the incident laser. Quantum features like dynamical back-action amplification, which are related to the vibrational modes and correlations between emitted photons of the molecules, can be described through the molecular optomechanical model~\cite{Schmidt2017,Aspelmeyer2014,M. k. Schmidt 2016,Schmidt2016}.

Several challenging phenomena such as distinguishing similar molecules and designing tunable optical switches have been studied in this framework~\cite{Liu2017}. Heat transfer of two adjacent molecules in a plasmonic nanocavity, optomechanical cooling in the nonlinear regime, collective effects in SERS, and optical mass sensing are other noticeable phenomena that have recently been investigated through this optomechanical model~\cite{Yin2011,Liu2017,Ashrafi2019,Dezfouli2019,Zhang2020}.

Liu et al. proposed recently a mass sensing setup with high precision at room temperature using an artificial molecule with a small effective mass instead of a real Raman active molecule in the vicinity of plasmonic nanostructures~\cite{Liu2018}. It is an interesting mass sensing system due to the large mass sensitivity of the mechanical resonator resulting from the large plasmon-phonon coupling~\cite{Liu2018,Li2013,Chen2014}.

The canonical model of cavity optomechanics can provide the ability to analyze the molecular plasmonic systems and subsequently investigate the effects of several parameters such as incident pump field, coupling strength, and plasmonic damping rate on the transmission spectrum. However, the explicit dependence of these parameters upon the material and geometrical features of the system has been studied less analytically~\cite{Dezfouli2017}. Establishing a formalism that prepares the physical ground to understand the influences of these parameters on the dynamics of the system and control the transmission peak height and shape of the probe beam, would be practically useful. Therefore, as the main purpose of this study, we intend to enter these features into the optomechanical formalism of the molecular plasmonic system.

Unlike the traditional method based on the first-principle approach to obtain polarizability of plasmonic nanostructure for sensing purpose~\cite{Yin2011}, in the present contribution, we start with the mode-selective quantization scheme to extract the optomechanical Hamiltonian describing the interaction between a plasmonic nanostructure and a suspended graphene nanoribbon (SGNR)~\cite{Dzsotjan2016,Castellini2018}.
This Hamiltonian contains the electromagnetic Green's tensor of the system through which dispersive and dissipative properties of plasmonic cavities enter into the formalism. We then focus on two types of isotropic and anisotropic spherical plasmonic cavities and derive the multipolar polarizability from the Green's tensor of the system in the quasi-static approximation.
In this way, we can analytically derive the optomechanical coupling strength, which includes the modal volume of the plasmonic subsystem. The modal volume is a crucial concept in the field of PCQED that quantifies the magnitude of the electric field confinement. Although this parameter is not the geometrical volume but depends on the geometrical properties of the nanocavities~\cite{Kristensen2014,Huang2016}.

This paper is organized as follows. In section II. we present the details of the mode-selective quantization method and then generalize this theoretical treatment to formulate a multimode optomechanical Hamiltonian expressing the interaction between plasmonic nanostructures and the SGNR. To further illustrate the flexibility of this method, in section III we determine the explicit expression of the optomechanical strength and modal volume for coupling of the isotropic and anisotropic spherical plasmonic cavities with the SGNR. We subsequently obtain the transmission spectrum of the probe field related to the two aforementioned plasmonic systems for the sensing process at room temperature. In section IV, numerical results associated with the spectral function, optomechanical strength, and transmission spectrum are depicted and discussed. The paper is summarized by exploring optimal mass sensing in section V. Details related to the mode-selective quantization, the quantum Langevin equations, and the Mie coefficients can be found in Appendices A to D, respectively.

\section{Theoretical framework}\label{Sec:Theoretical framework: A model and  basic relations}
\begin{figure}[h]
\includegraphics[width=1\columnwidth]{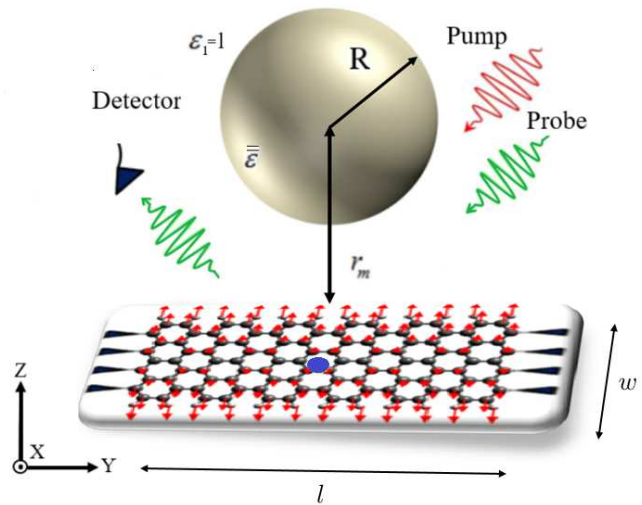}
\caption{(a) A schematic representation of the parametric interaction between the plasmonic cavity
mode and  suspended graphene nanoribbon in the presence of a strong
pump field and a weak probe field. Here, plasmonic nanostructure is placed
in vacuum. Double clamped graphene nanoribbon (armchair configuration) is located in the
$xy$ plane with periodicity direction along $y$.}\label{Fig.1}
\end{figure}

Consider a system (Fig.~\ref{Fig.1}) composed of a double clamped graphene
nanoribbon in the vicinity of a spherical plasmonic nanocavity with radius $R$ in the presence of
external fields. The graphene nanoribbon lies in the $xy$ plane, and its
axis direction is along $y$. We suppose that the nanoribbon is located
at the equilibrium distance $r_m$ from the center of the spherical cavity at
origin and can vibrate in horizontal direction. Here, the nanoribbon vibrations associated with the breathing-like mode, where all the atoms of the nanoribbon move in-plane along the ribbon width direction, as shown in Fig.~\ref{Fig.1}. This mode is Raman active due to the inversion symmetry of the atoms-displacement pattern in the nanoribbon~\cite{Gillen2010}. We can well characterize the vibrational mode of the SGNR by a harmonic oscillator with frequency $\omega _m$.
The relative permittivity of the background medium is taken to be that of free space, $\varepsilon_1=1$, and $\bar{\bar{\varepsilon}}$ is the relative permittivity tensor of the spherical nanocavity.

Free Hamiltonian of the whole system is given by $\hat H_{free} =
\hat H_m + \hat H_p$, in which the first term refers to the mechanical
vibrations of the SGNR, {\rm i.e.} $\hat H_{m} =
\hbar {\omega _m}{\hat b^\dag }\hat b$.
Here, $\hat b^\dag (\hat b)$ is the creation (annihilation) operator of the vibrational mode of the SGNR with the frequency $\omega _m$. $\hat H_{p}$ is the free Hamiltonian of the plasmonic subsystem and
describes the total energy of the electromagnetic field in a lossy
plasmonic medium, so it can be written as:
\begin{eqnarray}\label{free field Hamiltonian for plasmonic subsystem}
\hat H_{p} &= &\int {{d^3}r\int_0^\infty d\omega \, \hbar \omega \, {\bf{\hat f}^\dag }( {\bf r},\omega)\cdot{\bf{\hat f}} ({\bf r},\omega)},
\end{eqnarray}
where the bosonic operators $ \bf{\hat f}^{\dag}({\bf r},\omega) $ and $ \bf{\hat f}({\bf r},\omega) $  represent the collective excitations of the
electromagnetic field and the medium with the commutation relation $
\left[{\hat f}_i({\bf r},\omega),{{\hat
f}^\dag}_j ({\bf r}',\omega ') \right]= \delta _{ij}\delta
({\bf r} - {\bf r}')\delta (\omega  - \omega ')$~\cite{Gruner1996,Dung2000}.

Let us theoretically model the parametric coupling between
the optical modes of plasmonic structure and the vibrational mode of the SGNR in the framework of the molecular
optomechanics~\cite{Schmidt2017}. This interaction Hamiltonian can be described by the dipole interaction:
\begin{eqnarray}\label{interaction Hamiltonian}
\hat H_{int} =  - \hat {\bf P}({\bf r }_m)\cdot \hat{\bf E}({\bf r}_m).
\end{eqnarray}
Here, the induced Raman polarization $ \hat {\bf P}({\bf r}_m)$
can be identified with the Raman polarizability tensor of the
vibrational mode of the SGNR, $\bar{\bar\alpha }_R$,  as $ \hat
{\bf P}({\bf r}_m)=-\bar{\bar\alpha}_R\cdot
\hat{\bf E}({\bf r}_m)$. The Raman polarizability can be expressed in terms of
the quantized displacement operator of nanoribbon via the relation
$\bar{\bar\alpha}_R = - \sqrt {\hbar /2\omega _m} (\hat b +
\hat b^\dag)\bar{\bar {\cal R}} $, wherein the second rank Raman tensor $\bar{\bar {\cal
R}}$ depends on the molecular structure and bonds nature of the SGNR~\cite{Roelli2016}.

Based on the canonical quantization scheme for the electromagnetic
field in a dispersive and lossy medium~\cite{Gruner1996,Dung2000,Behbahani2016,Kheirandish2009}, the
positive frequency part of the electric field operator
$\hat{\bf E}^{(+)}$ can be written as
\begin{eqnarray}\label{component of electric field in term of Green tensor}
\hat{\bf E}^ {(+)} ({\bf r}_m,\omega ) &=& i(\omega^2/c^2)\int {{d^3}r\sqrt {\frac{\hbar {{\bar{\bar \varepsilon}} _I({\bf r},\omega )}}{\pi \varepsilon _0}}} \bar{\bar{{\bf G}}}({\bf r}_m,{\bf r},\omega )\nonumber \\
&& \cdot \bf{\hat f}(\bf{r},\omega ),
\end{eqnarray}
where $\bar{\bar \varepsilon}_I(\bf{r},\omega )$ is the
imaginary part of the permittivity tensor of medium, and
$\bar{\bar{{\bf G}}}({\bf r}_m,{\bf r},\omega )$ is the
classical Green's tensor satisfying the Helmholtz equation together
with appropriate boundary conditions. The symmetry-related considerations in the point group theory for the graphene nanoribbon~\cite{R. Gillen2010,Saito2010} along with the method of mode-selective quantization~\cite{Dzsotjan2016}
allow us to derive a final discrete form of the Hamiltonian based on the new frequency-independent plasmonic operators
$\hat a_n(r_m)$ at the position of nanoribbon as
\begin{eqnarray}\label{system Hamiltonian}
\hat H_{sys}&=&\sum\limits_{n = 1}^N {\hbar {\omega _n}\hat a_n^\dag ({\bf r}_m){{\hat a}_n} ({\bf r}_m)}  + \hbar {\omega _m}{\hat b^\dag }\hat b\\
&& - 2\hbar (\hat b + \hat b^\dag )\sum\limits_{n = 1}^N {g_{op,n}\hat a_n^\dag  ({\bf r}_m){\hat a_n} ({\bf r}_m)}\nonumber,
\end{eqnarray}
where, $\omega _{n}$ describes the $n^{th}$ resonance frequency
of the plasmonic cavity. Furthermore, the frequency-independent coupling
function $g_{op,n} = 2\left| {g_n}(r_m) \right|^2 $, which is known
as optomechanical coupling strength, can be obtained employing the spectral functions through the relation  $\left|
{g_n}(r_m) \right|^2 = \int {d\omega {\left|
k_n(r_m,\omega ) \right|}^2} $.  More details on how this optomechanical coupling spectra, $K_n={{\left| {k_n(r_m,\omega )} \right|}^2}$, is generated and also the steps for obtaining this Hamiltonian are outlined in Appendices A and B.

Mass sensing in the pump-probe technique requires examining the probe response of the SGNR~\cite{Yan2018,Gao2016}. To this purpose, the plasmonic subsystem is coherently driven by a strong pump field and a much weaker probe field, as shown in Fig.~\ref{Fig.1}. Given this, we can obtain the total Hamiltonian of the system only by adding driving terms to the Hamiltonian $\hat H_{sys}$. For more convenience, we recast the total Hamiltonian in a reference frame rotating at the pump frequency. It yields
\begin{eqnarray}\label{total Hamiltonian}
\hat H_{tot}& =& \hbar \omega _m \, {\hat b^\dag }\hat b + \sum\nolimits_{n = 1}^N {\left[ \hbar \Delta _n \, \hat a_n^\dag  ({\bf r}_m) \hat a_n ({\bf r}_m) \right.} \nonumber\\
&&\  - \hbar g_{op,n} \, \hat a_n^\dag  ({\bf r}_m) \hat a_n ({\bf r}_m)(\hat b^\dag  + \hat b)\nonumber\\
&&\ - i\hbar \Omega _{pr}\left( \hat a_n ({\bf r}_m) e^{i\delta t} - \hat a_n^\dag  ({\bf r}_m)e^{- i\delta t} \right)\nonumber\\
&&\ \left.  - i\hbar \Omega _{pu} \left( \hat a_n ({\bf r}_m) - \hat a_n^\dag  ({\bf r}_m) \right) \right],
\end{eqnarray}
where $\Omega _{pu}$ and $\Omega _{pr}$ are coherent driving
coupling parameters related to the $n^{th}$ localized surface
plasmonic ($LS{P_n}$) mode with $\Omega _i = \kappa _n\sqrt {\varepsilon
_0 V_{eff}/2h \omega _n} E_{i,m}/2\,(i = pu,pr)$
~\cite{Esteban2014}. Here, $\kappa _n = \Gamma _n/2$ and
${E_{i,m}}$ are the damping rate of $LS{P_n}$ mode and the maximum near
field scattered by the structure, respectively. The $E_{i,m}$ can be written in
terms of plasmonic enhancement factor and incident pump field
~\cite{des Francs2016}. The parameter $\Delta_n= \omega_n -
\omega _{pu}$  denotes the detuning between the pump and the
$n^{th}$ mode plasmonic frequency, and the detuning between the probe and
pump field is defined as $\delta  = \omega_{pr} -\omega_{pu}$.

Using the Heisenberg-Langevin approach~\cite{Scully1997,Aspelmeyer2014}, which includes dissipation and fluctuation mechanisms in the system, we can determine the time evolution of the plasmonic annihilation operator and the mean response of the mechanical subsystem. Since the incident probe field is much weaker than the pump field, we can employ the
perturbation method to investigate the response of the probe field.
Appendix C provides the derivation details of output amplitude in the frequency of probe field based on the pump-probe
technique and input-output theory. These calculations help us to build the output amplitude  $ a_{n + ,out}=
\sqrt {2\kappa _n} \, a_{n + }$, or equivalently, define the transmission of the probe beam as the ratio of the output and input amplitudes, {\rm i.e.} $t(\omega _{pr})=1-2\kappa _n  a_{n + }/\Omega _{pr}$
~\cite{Aspelmeyer2014}. We will show that the $ a_{n
+}$ contains all information about optomechanical strength and
plasmonic damping rate of the $n^{th}$  plasmonic mode as well as
the material and geometrical features, as will be seen in Eq. (C4). Determining the coupling strength
in terms of the Green tensor and using the procedure presented in the next section, we would
analyze the probe response associated with the several plasmonic subsystems.
%
\section{plasmonic cavities}
To illustrate how geometrical and material features of plasmonic structure can modify the dynamic of the system and consequently the mass sensing precision, in the present section, we consider two cases and determine the explicit form of the optomechanical strength for them. We then study probe responses in each case.

\subsection{Anisotropic spherical nanostructure optomechanically coupled to the SGNR}
In the first case, we consider a uniaxial anisotropic sphere illuminated by a plane wave. The form of the
relative permittivity tensor for this anisotropic sphere is given by
\begin{eqnarray}\label{permitivity tensor }
\bar{\bar{\varepsilon}}=  \left[ (\varepsilon _r - \varepsilon _t){\bf \hat r} {\bf \hat r} + \varepsilon _t\bar{\bar{\bf I}} \right],
\end{eqnarray}
where $\bar{\bar{\bf I}}$ is the unit dyad, and $\varepsilon _r$ and $\varepsilon _t$ are the radial and
tangential components of the permittivity tensor, respectively
~\cite{Qiu2007}. With the general form of Green's tensor in
hand and applying the symmetry consideration related
to the radial anisotropic sphere, the scattering part of the Green's
tensor for the tangential direction is simplified as follows
  \begin{eqnarray}\label{scattering Green tensor}
&&\bar{\bar{G}}_{s,tt}^{(11)}(r_m,r_m) = \frac{i \omega}{8\pi c }\sum\limits_{n = 0}^\infty  (2n+ 1)\\
&&\times \left[B_N^{11} {\Big(\zeta '_n ({k_1}{r_m})/{k_1}{r_m}\Big)^2} + B_M^{11} {\left( h_n^{(1)}({k_1}{r_m}) \right)^2} \right],\nonumber
\end{eqnarray}
where $k_1=\omega/c$, $h_n^{(1)}({k_1}{r_m})$ is the first-type spherical Hankel
function in the position of the nanoribbon, $\zeta _n({k_1}{r_m}) =
({k_1}{r_m}) h_n^{(1)}({k_1}{r_m})$ is the spherical Riccati-Hankel function of the first kind, and the primed function refers to the derivative with respect to its argument. For uniaxial anisotropic sphere with source and field points in the first layer, anisotropy effects appear only in the Mie coefficients $B_{l}^{(11)}=  -
T_{l,12}^{(1)}/T_{l,11}^{(1)}$ with $l=N,M$. Here, $T_{l,12}^{(1)}$
and $T_{l,11}^{(1)}$ are the elements of transmission T-matrix
that are derived in Appendix D. If  the radius of the sphere is very
 small compared to the wavelength of the incident
field, we can restrict our attention to the QSA and simplify the Mie
coefficients in this limit~\cite{Abramowitz1972,Bohren1998,Des
Francs2009}.

As described in Appendix D, the relevant Mie coefficient in our system is
$B_N^{(11)}$. It is shown that in the QSA, $B_N^{(11)}$
will be proportional to the modified quasi-static polarizability.
This modified polarizability takes exactly the same form of the
polarizability as an isotropic sphere if we define an effective
permittivity $\varepsilon _{eff}^{ani}$ as below
 \begin{eqnarray}\label{permitivity of anisotropic sphere }
\varepsilon _{eff}^{ani} =\frac{\upsilon}{n}{\varepsilon _{r}}.
\end{eqnarray}
Here, $\upsilon = \left[
n(n + 1){AR + 1/4} \right]^{1/2} - 1/2$, in which the anisotropy ratio $AR$ defined as $AR
= \varepsilon _t/\varepsilon _r$~\cite{Qiu2007}. In what
follows, we assume both radial and tangential components of the
permittivity tensor in Eq. (8) are given by the Drude-like model
$\varepsilon _i = \varepsilon _{\infty i} - \omega
_{pi}^2/\omega (\omega  + i \Gamma _{pi})$ with $i = t,r$. We also suppose that the optical parameters of two
components obey the relations $\omega _{pr}^2/\varepsilon _{\infty r} = \omega
_{pt}^2/\varepsilon _{\infty t}$ and $\Gamma _{pr} =\Gamma
_{pt}=\Gamma _{p} $. According to these assumptions, the effective permittivity
$\varepsilon _{eff}^{ani}$  takes a Drude-like form,
 $\varepsilon _{eff}^{ani} =\varepsilon _\infty ^{ani} - (\omega_{p}^{ani})^2/\omega (\omega  + i \Gamma _{p})$, in which
the optical parameters are defined as follows:
\begin{subequations}\label{Drude-like parameters}
\begin{eqnarray}
\omega _p^{ani} &=&\left( {\left( n(n + 1)A{R_{\infty }}/{n^2} + 1/4{n^2} \right)}^{1/2} - 1/2n \right)^{1/2} \omega _{pr},\nonumber\\
\\
\varepsilon _\infty ^{ani}& =&\left( {\left(n(n + 1)A{R_{\infty }}/{n^2} + 1/4{n^2} \right)^{1/2}} - 1/2n \right) \varepsilon _{\infty r}.\nonumber\\
\end{eqnarray}
\end{subequations}
Here, $\omega _p^{ani}$ and $\varepsilon _\infty ^{ani}$
represent the effective bulk plasmon frequency and the effective
high-frequency limit of the dielectric function, respectively. Furthermore, the anisotropy ratio reduces to the constant
parameter $A{R_{\infty }} = {\varepsilon _{\infty t}}/{\varepsilon _{\infty
r}}$ ~\cite{Des Francs2009,Des Francs2011}. Therefore, we can easily find the
approximate imaginary part of the Green tensor in the tangential
direction near the resonance frequencies of the plasmonic cavity
(for more details refer to Appendix D). Substituting this result into Eq. (A6), we get the approximate expression for the near-field frequency-dependent optomechanical coupling spectra as:
\begin{eqnarray}\label{spectrul function }
K_n &\approx & \sqrt {\frac{\hbar }{2 \omega _m}} \left( {\frac{\bar {\cal R} \, {\omega _n^{ani}}}{16\pi\varepsilon _{0}}} \right)\frac{{n(n + 1)(2n + 1) R^{2n + 1}}}{[n \, \varepsilon _\infty ^{ani} + (n + 1)]r_m^{2n + 4}}\nonumber\\
&&\times \frac{\Gamma _n^{ani}/2\pi }{{(\omega _n^{ani} - \omega )}^2 + {(\Gamma _n^{ani}/2)^2}} .
\end{eqnarray}
Here, $R$, $r_m$ and $\bar {\cal R}$
represent the radius of the anisotropic sphere, the separation
distance of the SGNR from the center of sphere and
diagonal elements of the Raman tensor ${\cal R}_{ii}$ for $i
=x$, respectively. The resonance frequency $\omega _n^{ani}$ and the total width $\Gamma
_n^{ani}$ corresponding to the $n^{th}$ localized surface
plasmonic mode ($LS{P_n}$) of the anisotropic sphere (effective
sphere) are given in Appendix D. The coupling spectra $K_n$ in Eq.
(10) can be exploited to extract the modified optomechanical coupling
strength of the anisotropic sphere as
\begin{eqnarray}\label{optomechanical coupling }
g_{op,n}^{ani} \approx \sqrt {\frac{\hbar }{2 \omega _m}} \left({\bar {\cal R} \, \omega _n^{ani}}\right)\frac{n(n + 1)(2n + 1){R^{2n + 1}}}{8\pi\varepsilon _{0} (n \, \varepsilon _\infty ^{ani} + (n + 1))r_m^{2n + 4}}.\nonumber\\
\end{eqnarray}
By defining the mode volume of the plasmonic nano-cavity (anisotropic sphere) for $n^{th}$ localized surface plasmonic mode as
\begin{eqnarray}\label{modal volume }
V_n^{ani} = \frac{8\pi (n \, \varepsilon _\infty ^{ani} + (n + 1))r_m^{2n + 4}}{n(n + 1)(2n + 1) R^{2n + 1}},
\end{eqnarray}
the following expression for the parameter, $g_{op,n}^{ani}$, which introduced in Ref.~\cite{Yin2011}, is
recovered:
\begin{eqnarray}\label{final opromechanical coupling }
g_{op,n}^{ani} = \bar {\cal R}\sqrt {\hbar /2{\omega _m}} {\rm{ (}}\omega _n^{ani}/\varepsilon _{0}V_n^{ani}{\rm{)}}.
\end{eqnarray}
%
\subsection{Isotropic spherical nanostructure optomechanically coupled to the SGNR}
To demonstrate the role of anisotropy in the dynamics of the system and provide a quantitative comparison with its isotropic counterpart, we explore results for the second case of an isotropic spherical cavity, i.e., the limiting case  $\varepsilon _t = \varepsilon _r$.

In this limit, the modified Mie coefficients that derived in Eqs.
(D2a) and (D2b) reduce to the coefficients for the
isotropic sphere, as developed in Ref.~\cite{Li1994}. Furthermore,
the permittivity function in Eq. (8) simplified to the
expression with Drude parameters $\omega _p$ and $\varepsilon
_\infty$ for the isotropic sphere. So, we can identify the optomechanical
strength and modal volume for isotropic sphere by replacing the Drude-like
parameters $\omega _n^{ani}$ and $\varepsilon _\infty ^{ani}$ with
their limiting parameters $\omega _n$ and $\varepsilon _\infty $  in
Eqs. (11) and (12). In this way, the results in Eqs.
(11) and (12) for $A{R_{\infty }} = 1$ properly reduce to the corresponding expressions for the isotropic
sphere. In section IV, we depict and compare the effects
of these physical features on the sensing process for several plasmonic
structures.


\section{RESULT AND DISCUSSIONS}\label{Sec:Entanglement}
\subsection{Optomechanical coupling Strength}
In this section, we numerically study the behavior of the optomechanical coupling spectra and strengths for interaction between each plasmonic nanostructure and the SGNR. We compare the obtained results for different plasmonic structures. We then explore the effects of the material, geometrical and modal parameters on the system dynamics and consequently the sensing process.

First of all, we list the physical parameters of the subsystems.
Consider a graphene armchair nanoribbon of dimensions $l =20nm$ and $w\approx 20 nm$ with the frequency of the breathing-like mode $\omega _m\approx 470GHz$~\cite{Liu2020,Zhou2008}, and the total mass $m
\approx 3 \times {10^{ - 22}}kg$~\cite{Hiura2012,Kalosakas2021}.
The quality factor of the Raman active mode drastically decreases at room
temperature. This results in a damping rate of $\gamma  =
1.9GHz$. We set the quantum yield of the nanoribbon $\eta  = 0.01$ and assume
that the square of the Raman tensor element $\bar {\cal R}$ is of order
$10^3 ({A^4}a.m.{u^{ - 1}})$~\cite{Roelli2016}.
For the isotropic nanocavity, we take into account a silver nanosphere of radius $R =
10nm$ placed at a distance of $r_m= 14nm$ from the graphene nanoribbon, and characterize their dissipative
and dispersive properties by the Drude model with typical parameters: $\omega _p = 1.9PHz$, $\Gamma _p=0.012PHz$, and $\varepsilon _\infty= 6$~\cite{Hakami2014,Van Vlack2012,Johnson1972}.
For the anisotropic nanocavity in which both radial and tangential components of the permittivity tensor described by the Drude-like model, we also choose the above material parameters for the radial component along with a variable anisotropy ratio. Using these parameters, and Eqs. (9a) and (9b), we can easily get the effective material parameters $\varepsilon _\infty ^{ani}$ and $\omega _p ^{ani}$.
\begin{figure}[h]
\includegraphics[width=1\columnwidth]{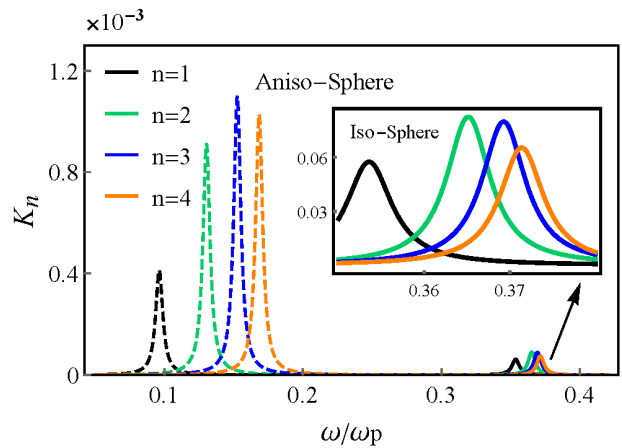}
\caption{Optomechanical coupling spectra versus $\omega /{\omega
_p}$ for coupling of the SGNR to the four first  $LS{P_n}$ modes of the
silver sphere (solid curves) and the anisotropic nanosphere (Dashed curves )
 with $AR_{\infty}= 0.01$. Here, we set $r_m= 14 nm$ and $R = 10 nm$.
Other material parameters are chosen as $\omega_p =  0.19 PHz$, $\omega_m= 470GHz$ and
$\gamma  = 1.9GHz$. The inset shows the zoomed-in of the spectra for the silver sphere nanocavity.}\label{Fig:2}
\end{figure}
\begin{figure}[h]
\begin{subfigure}[h]{1\linewidth}
\centering
\includegraphics[width=.97\columnwidth]{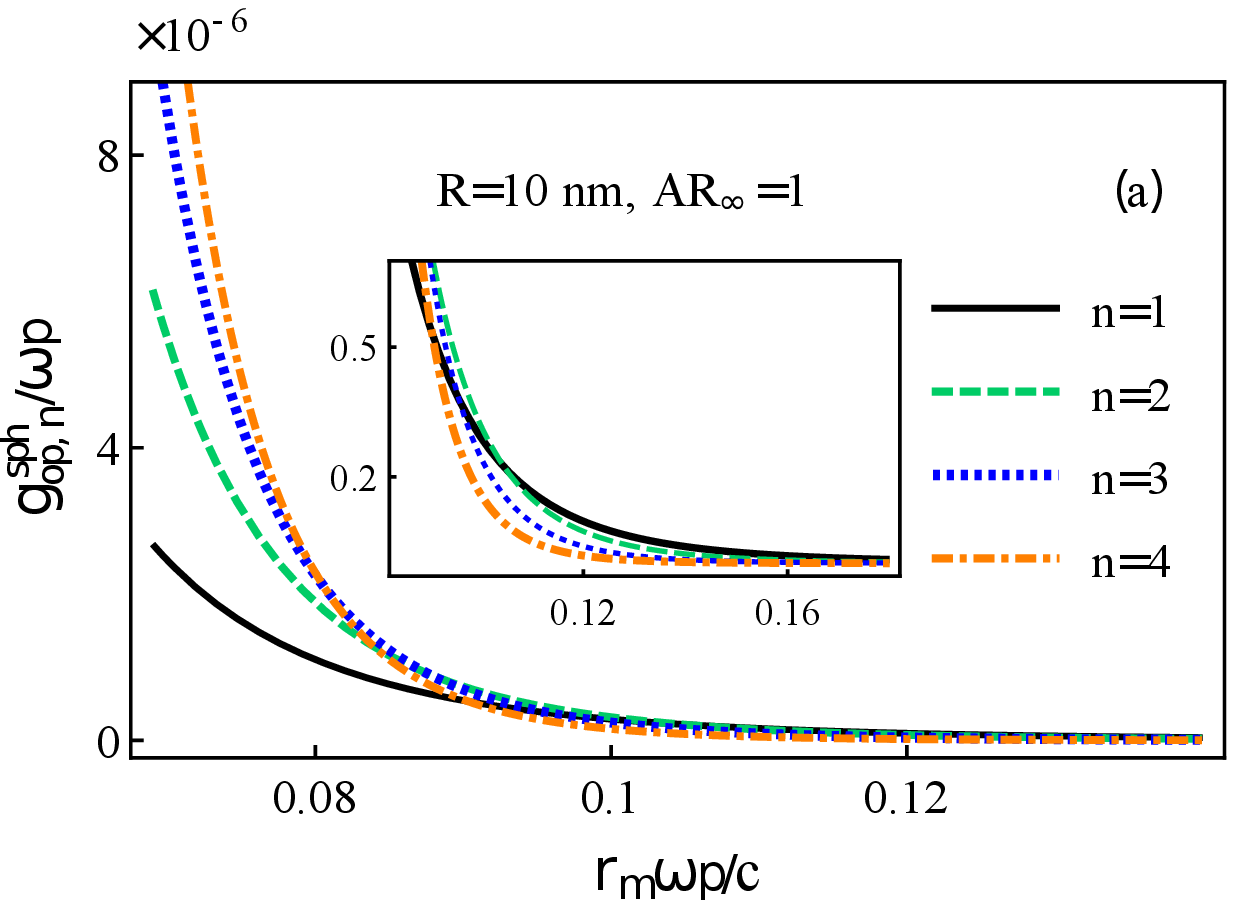}
\end{subfigure}
\\
\begin{subfigure}[h]{1\linewidth}
\centering
\includegraphics[width=.97\columnwidth]{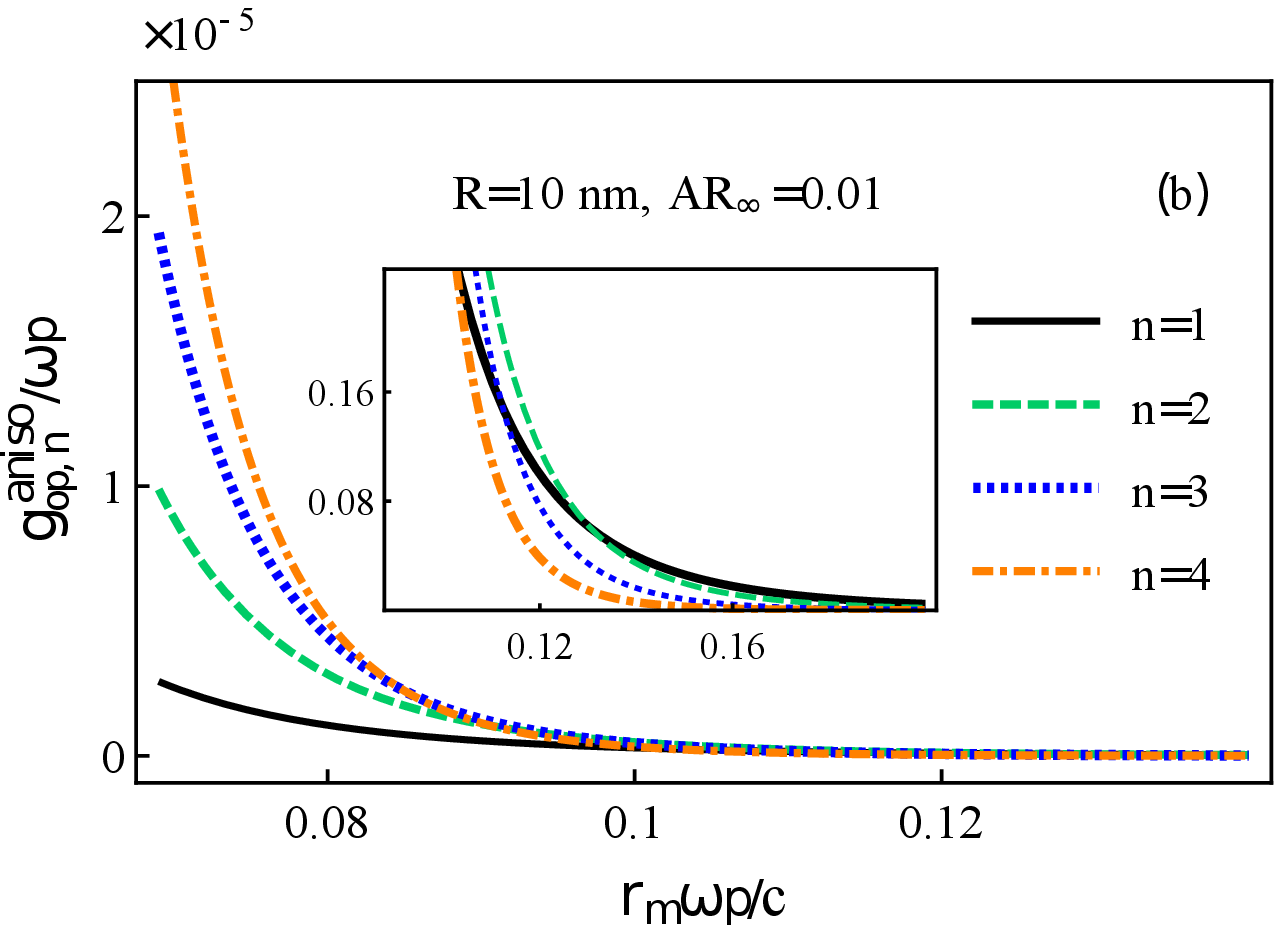}
\end{subfigure}
\caption{ Optomechanical coupling strength versus $r_m{\omega _p}/c$
 for interaction between the SGNR and four first  $LS{P_n}$ modes of (a) the
silver nanosphere, and (b) the anisotropic nanosphere.
 Values for parameters are the same as those in Fig.~\ref{Fig:2}.
 The insets show the zoomed-in of the panels within the range
 $r_m\approx0.084{c}/{\omega _p}$ to  $r_m\approx0.19{c}/{\omega _p}$.
Scaled separation distances are approximately equivalent to the interval distance $r_m=11 nm-22 nm$.}\label{Fig:3}
\end{figure}

We can now analyze the coupling of the SGNR to the plasmonic
modes of the nanocavity.
The solid curves in Fig. 2 represent the Lorentzian
coupling spectra $K_n$  for coupling of the four first $LS{P_n}$
modes of the silver nanosphere to the SGNR as a function of a
dimensionless frequency $\omega /{\omega _p}$.
We observe that the Lorentzian peaks associated with the spectral coupling functions $K_n$ start to overlap each other by increasing the order of multipolar coupling $n$. Thus, it seems the silver sphere nanocavity is not a suitable structure for mass sensing since spectral functions can not be resolved properly and overlap noticeably, especially for higher mode couplings.

Dashed curves in Fig. 2 are related to the Lorentzian peaks for an anisotropic spherical nanocavity manifesting the role of anisotropy in the dynamics of the system.
As it is evident, spectral
functions are slightly red-shifted and narrowed by decreasing the anisotropy ratio from $AR_{\infty}= 1$ to $0.01$.
Furthermore, there is no significant spectral interference compared to the isotropic nanosphere case. Therefore, by manipulating the anisotropy ratio, the Lorentzian curves are split from each other, and the position of peaks can be tuned. All of these features become practically useful in the sensing process and will be addressed in more detail later.

The optomechanical coupling strength corresponding to the interaction between the SGNR and several multipolar modes of the isotropic and anisotropic nanospheres are illustrated as a function of the dimensionless distance $r_m{\omega _p}/c$ in Figs. 3(a) and 3(b), respectively.
One clearly sees that the largest values for
the coupling strength belong to the interaction between the plasmonic modes of the anisotropic
spherical nanocavity and the SGNR,
regardless of the magnitude of $n$ in Fig. 3 (b).
Comparing the behavior of the several multipolar couplings in two panels, we find that for separation distances up to
$r_m \approx 19nm$, coupling of the SGNR to the higher-order modes plays an
essential role in the dynamics of the system. However, the dipolar coupling, $n=1$, becomes dominant for large distances.
 As expected, the optomechanical coupling strength is greater for small separation distances in Figs. 3 (a) and 3 (b).

\subsection{Mass Sensing}
Having explained the behavior of the coupling strength, we now present numerical results for mass
sensing through the pump-probe technique~\cite{Li2009,He2010,Jun2015}. Our purpose is to manipulate the geometrical
and material features of the spherical nanostructures to
see how the probe spectrum is effected at room temperature. Using
Eqs. (C4) and (C5), we can obatin the transmission spectrum of the probe field.
In Fig. 4(a), we have illustrated the probe fields transmission spectra for the
optomechanical couplings of the isotropic and anisotropic nanospheres
to the SGNR in terms of the probe-cavity detuning $\delta$
at fixed separation distance $r_m= 14nm$. Notice that here
$\delta$ is specifically defined as $\delta = \omega_1 - \omega_{pr}$,
which corresponds to the dipolar coupling.
\begin{figure}[h]
\begin{subfigure}[h]{1\linewidth}
\centering
\includegraphics[width=1\columnwidth]{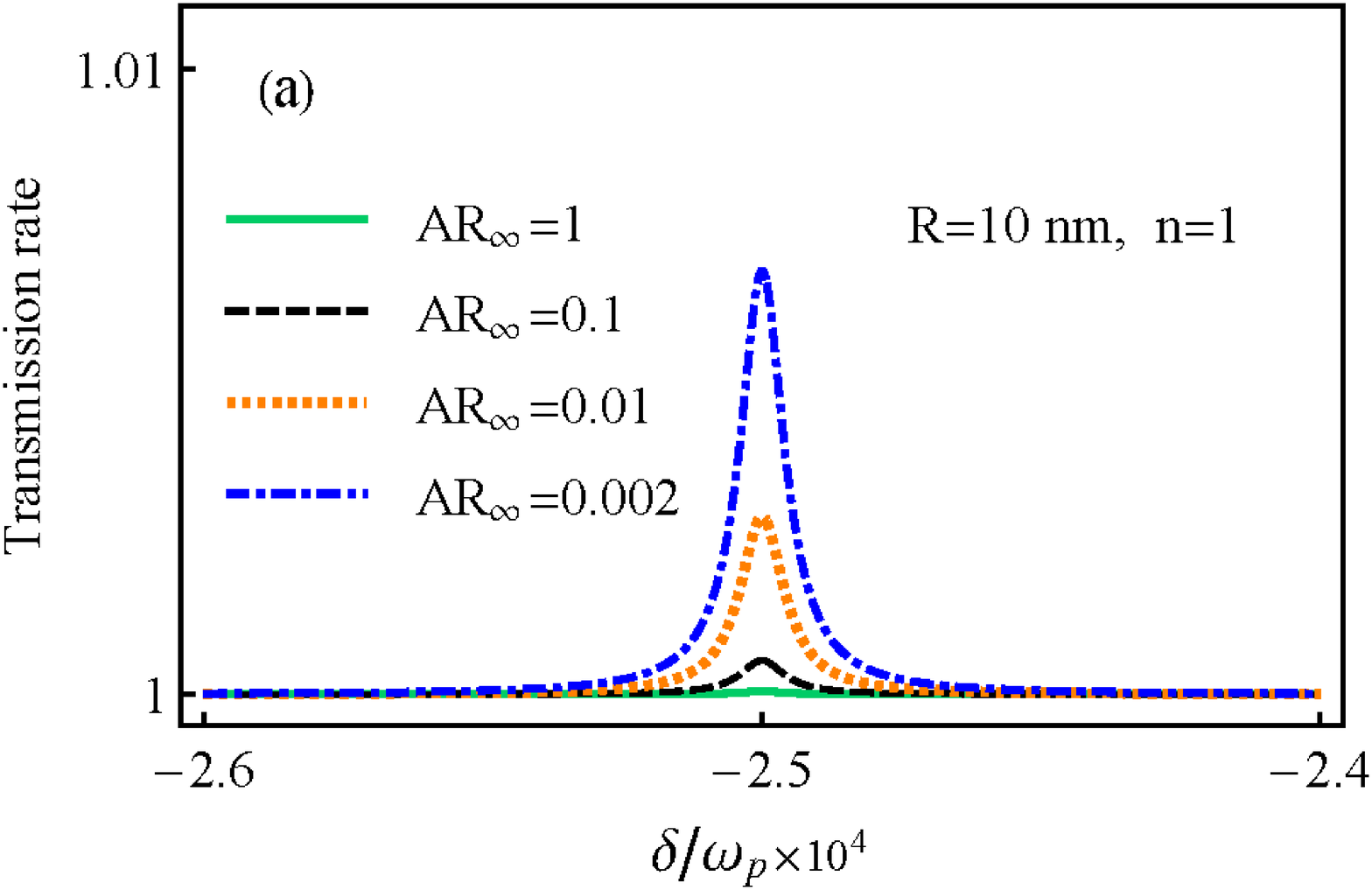}
\end{subfigure}
\\
\begin{subfigure}[h]{1\linewidth}
\centering
\includegraphics[width=1\columnwidth]{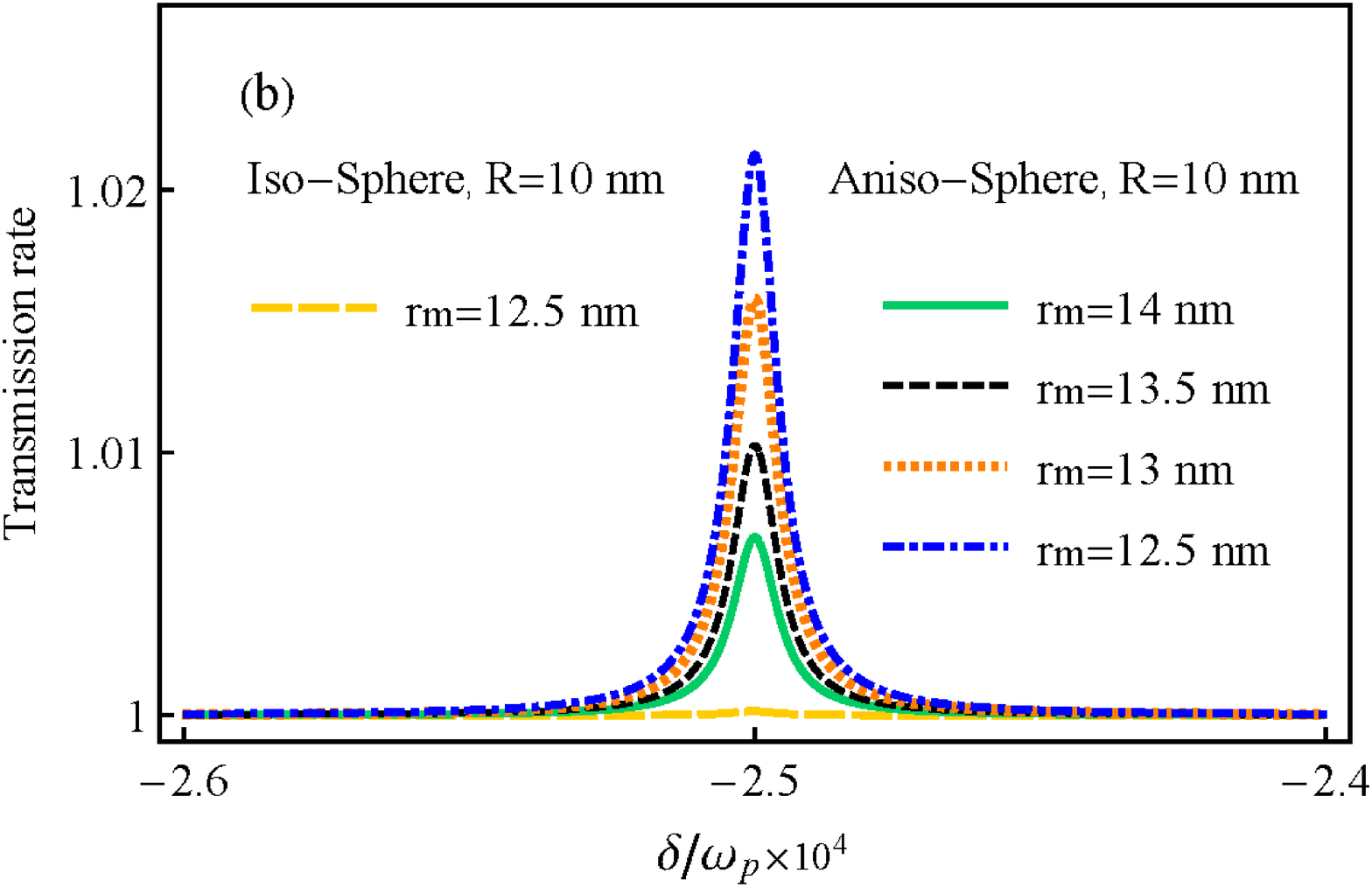}
\end{subfigure}
\caption{Transmission rate versus $\delta/{\omega _p}$ for dipolar coupling
of the SGNR to (a) the silver nanocavity, and (b) the anisotropic nanosphere with $AR_{\infty }=0.002$.
Values for other parameters are the same as those in Fig.~\ref{Fig:2}.}\label{Fig:4}
\end{figure}

We analyze the results for $AR_{\infty }=0.1$ (black dashed curve), $AR_{\infty }=0.01$ (orang dotted curve) and $AR_{\infty }=0.002$ (blue dot-dashed curve).
It is seen that the strength of the transmission peak enhances by decreasing  $AR_{\infty }$ from $ 1$ for the isotropic silver nanosphere (green curve) to the smaller value for the anisotropic nanosphere.
It is related to the role of mode volume as well as damping rate of
plasmonic nanocavities. Traditional CQED investigates the dynamics of atom-field interaction by exploring the influences of the cavity quality factor, while in the PCQED the key parameter is mode volume ~\cite{Ginzburg2016,Scully1997}. Plasmonic nanocavities confine electromagnetic energy beyond the classical diffraction limit, resulting in small mode volumes. Since coupling strength is inversely proportional to the mode volume, the nanostructures with smaller mode volume provide larger coupling strength and consequently higher transmission peak. When the anisotropy ratio decreases from the value $AR_{\infty }=1$ to $AR_{\infty }=0.01$, the enhancement of the optomechanical coupling strength occurs due to mode volume reduction.
This is verified by the plots in Figs. 3(a) and 3(b), in which the optomechanical coupling strength of the SGNR to the anisotropic nanocavity modes with $AR_{\infty }=0.01$ is stronger than the silver sphere modes regardless of the order of interaction.

It is worth noting that further reduction of the anisotropy ratio leads to the smaller optomechanical coupling strengths compared to the value of $0.01$   (here the figure is not presented for the sake of brevity). But on the other hand, the redshift of the resonance frequencies leads to the enhancement of the coupling functions between the plasmonic field and the classical incident fields ($\Omega _{pu}, \Omega _{pr}$).  These features provide a higher transmission peak for the smaller anisotropy ratio, as shown in Fig. 4(a).

 For nanostructures with a small radius, plasmonic damping rate comprises two terms: the Joule term that
accounts for electron scattering losses, and the radiative damping term~\cite{Bohren1998,Des Francs2009}. Having estimated the damping rates related to the dipolar mode of the two plasmonic nanocavities, we find that the total damping rate of the anisotropic nanosphere becomes slightly smaller than the damping rate of the silver sphere nanocavity due to the smaller contribution of the radiative damping term.

In Fig. 4(b), we illustrate the influence of separation distance $r_m$ on the probe transmission spectrum for dipolar coupling of the anisotropic nanosphere to the SGNR. Here, the results are depicted for a fixed
anisotropy ratio $AR_{\infty }=0.002$. As expected, when the SGNR gets closer to the surface of the anisotropic nanosphere, the intensity of the transmission peak enhances slightly which makes it easier to detect for sensing purposes.
For the SGNR placed $4 nm$ away from the anisotropic nanosphere surface, the result is shown with the green-solid curve in Fig. 4(b). We observe that the peak intensity for the anisotropic nanosphere at a farther distance is greater than that for the isotropic sphere at a shorter distance (yellow-solid line). Notice that for distances less than $r_m = 11 nm$, nonlocal effects become increasingly pronounced, which are out of the scope of the present study.

So far, we have examined the effects of the anisotropy ratio and separation distance on the transmission spectrum. In Fig. 5(a), we compare the behavior of the spectrum peaks for the coupling of the SGNR to the three first $LS{P_n}$ modes of the anisotropic nanosphere. Here, we set  $AR_{\infty }=0.002$ and $r_m = 12.5 nm$.
Results indicate that close to the surface of spherical nanocavity, higher-order mode coupling exhibits a higher peak intensity.

The sensing process with a weak incident pump field provides a suitable platform for the exploration of living organisms' features. Therefore, in Fig. 5(b), we have depicted the transmission peak strength for the probe field in terms of the incident pump intensity of the order $I_{pu} \approx10 KW/c{m^2}$. The corresponding results for higher intensities are plotted in the inset of Fig. 5(b).
\begin{figure}[h]
\begin{subfigure}[h]{1\linewidth}
\centering
\includegraphics[width=.97\columnwidth]{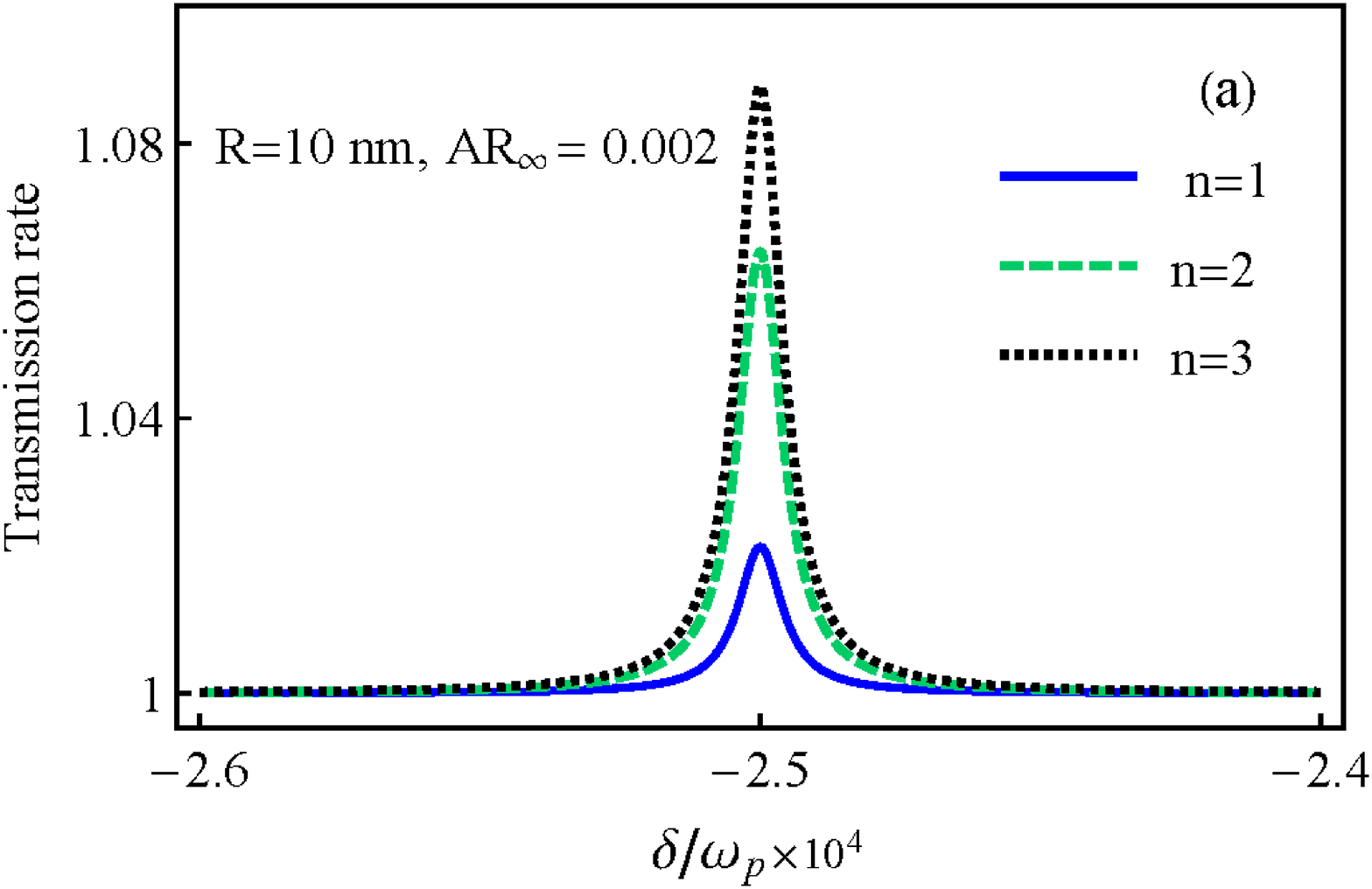}
\end{subfigure}
\\
\begin{subfigure}[h]{.98\linewidth}
\centering
\includegraphics[width=.93\columnwidth]{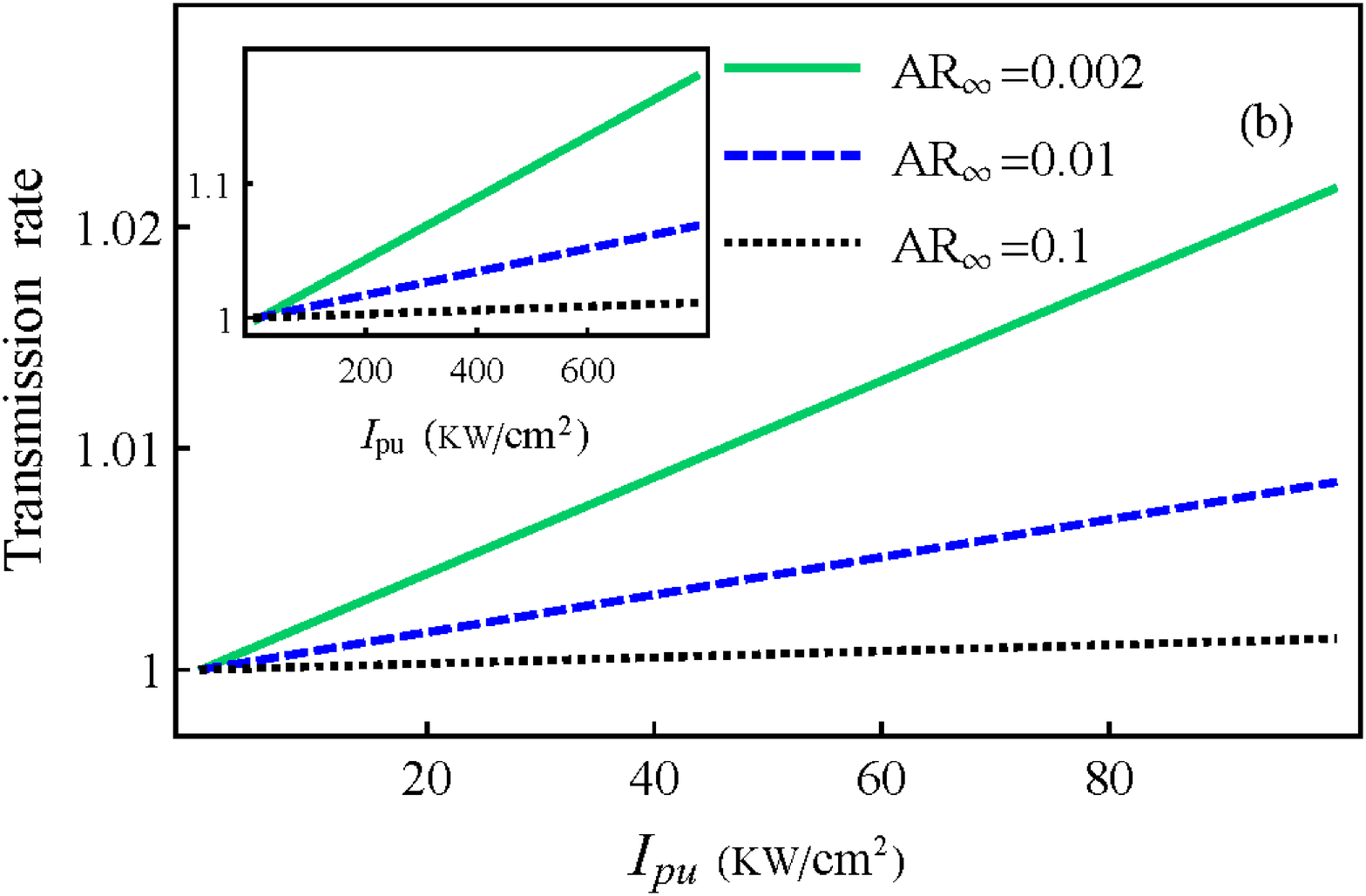}
\end{subfigure}
\caption{(a) Transmission rate versus $\delta/{\omega _p}$ for coupling the SGNR to
several multipolar modes of the anisotropic sphere with $R = 10nm$ and $r_m= 12.5nm$. (b)
Transmission rate versus incident pump intensity. Other
parameter values are $\omega _p =  0.19 PHz$, $\omega _m= 470GHz$ and
$\gamma  = 1.9GHz$. }\label{Fig:5}
\end{figure}

The extremely small size of the SGNR makes physical properties greatly sensitive to the perturbations originating from the external force and mass adsorbed on it. Based on the relationship between mass changes of the graphene nanoribbon and its frequency shift, one can perform a sensing process through the relation $\Delta m =(2m)\Delta \omega /{\omega _m}$~\cite{Liu2020}.
In the pump-probe scheme, the frequency shift induced by the molecular species loaded onto the nanoribbon surface is monitored. Minimal measurable mass depends on the bandwidth of the absorption spectrum. To understand the accuracy of mass sensing in this system, we make a numerical estimation in the following.

As it is evident from the black dot curve in Fig. 5(a), the octupolar coupling results in the highest peak for the transmission spectrum at the separation distance $r_m= 12.5nm$ when the anisotropic nanosphere is illuminated with the incident pump intensity $I_{pu} \approx400 KW/{cm^2}$. Therefore, we can estimate FWHM (full width at half maximum) of the probe spectrum for the black dot curve in Fig. 5(a), which is $\Delta \omega \approx 0.18GHz$.
The calculation indicates that the minimum resolution for mass sensing takes the value $\Delta m \approx1.2 \times {10^{ - 24}} kg$. For lower intensities of order, $I_{pu} \approx 40 KW/{cm^2}$, the height of the transmission peak reduces to the approximate value $0.01$. However, the accuracy of mass sensing becomes almost the same as before.

Notice that the analytical method introduced here can be generalized to include a variety of plasmonic nanocavities as the optical subsystem.
For multilayered plasmonic nanostructures, optical properties can be adjusted more accurately due to the tunable geometrical and material parameters of these structures. This feature makes plasmonic nanocavities attractive for sensing applications since the optomechanical coupling strength, as well as the plasmonic damping rate, can be adjusted more effectively through manipulating optical features of these multilayered nanostructures.
Tunable plasmonic damping rate and coupling strength affect the width and height of the probe transmission spectrum, respectively. Thus, an anisotropic spherical nanoshell (core) in combination with the metallic core (nanoshell), which can provide mass sensing with higher precision, will be the subject of our next study.

In the end, the nanomechanical subsystem may be affected by the thermomechanical and momentum exchange noises, as well the Casimir force. The contribution of the first two effects is negligible at room temperature ( see Ref.~\cite{Ekinci2004}), while an approximate estimation of the Casimir force to examine its effect on the sensing process is needed. Bimonte et al. derived a semi-analytic formula, $F_c =- ({\pi ^3}\hbar cR/360{h^3})$, for the sphere-plate Casimir force, in which $R$ is the radius of the spherical structure, and $h$ represents the separation distance between the plate and surface of metallic sphere~\cite{Bimonte2017}. For parameter values  $R=10nm$ and $h=3nm$, the Casimir force takes the value $F_c = 9 \times {10^{ - 27}}N$.
In another study, Biehs et al. investigated the Casimir-Polder force between a gold nanoparticle and a single sheet of pristine graphene in detail~\cite{Biehs2014}. They have shown that although the graphene sheet behaves like a perfect metal for distances larger than the thermal wavelength, for small distances the Casimir-Polder force is relatively small compared to the values of the ideal metal case and depends on the Fermi level of the graphene sheet. Bearing this in mind, it is clear that the weight of the minimum measurable mass would be still comparable with the Casimir force.
%
\section{Summary and conclusion}\label{conclusion}
%
In this paper, we have investigated the anisotropy effects of nanocavity on mass sensing in the molecular plasmonic system. To do so, we have presented a formalism that allows us to manipulate the transmission spectrum of the probe field for mass sensing through the pump-probe technique. In this manner, the molecular plasmonic system is mapped onto the cavity optomechanical model~\cite{Roelli2016,Schmidt2017}, then the mode-selective quantization scheme presented in Ref.~\cite{Dzsotjan2016} is extended to formulate a multimode Hamiltonian of the optomechanical system.

It is shown that the geometrical, material, and modal considerations, which are relevant to the plasmonic subsystem, are contained in the Hamiltonian through the electromagnetic Green's tensor.
We found the explicit form of the optomechanical coupling strength, the parametric interaction between the plasmonic mode, and the Raman active mode of the SGNR in terms of the symmetrical and geometrical properties of the graphene nanoribbon and plasmonic nanocavity. Taking symmetry properties of the nanoribbon into account, we determined the coupling spectral function and optomechanical strength for some spherical plasmonic structures by extracting their Green tensors.
We showed that an anisotropic nanocavity with a small anisotropy ratio is more suitable for the sensing process. Furthermore, the manipulation of anisotropy features of the plasmonic nanocavity made it possible to reduce mode volume (enhance optomechanical strength) and damping rate for optimal mass sensing.
We then have illustrated transmission spectra for two cases of plasmonic nanostructures. The results indicated an enhancement in the transmission peak corresponding to the anisotropic nanocavity, and verified its superiority over the isotropic one for sensing applications.

In a traditional optomechanical system, a high-Q microcavity at cryogenic temperature is needed to accurately perform mass sensing. Recently, mass sensing with the precision of order $10^{ -26} kg$ is performed in molecular plasmonic systems with low-Q plasmonic nanocavities (cylindrical dimer) at room temperature~\cite{Liu2018}. Of course, we can attribute this high efficiency to the small mode volume of the plasmonic nanocavity, and as well a controversial frequency considered for the breathing-like mode of the graphene nanoribbon.
Unlike the previous work developed by Liu, as a main purpose of this paper, we
have presented an analytical method to explicitly calculate the mode volume and damping rate of plasmonic modes for different plasmonic nanocavities.
Of course, improving the accuracy of the current mass sensing device and achieving the highest accuracy by engineering the plasmonic multilayered nanostructures is under consideration.

\appendix
\section{Mode-selective quantization approach}\label{App:Green tensor}
%
The Green's tensor contains all modal information of the plasmonic
structure. Following the method developed in
Refs.~\cite{Dzsotjan2016,Castellini2018}, we consider a spherical geometry for which the Green's tensor can be decomposed as a sum over the discrete modes (multipole expansion with spherical symmetry):
\begin{eqnarray}\label{Green tensor decoposition}
{\bar{\bar{\bf G }}}({\bf r}_m, {\bf r},\omega ) =\sum\limits_{n = 1}^\infty  {\bar{\bar{\bf G}}}_n ({\bf r}_m, {\bf r},\omega ),
\end{eqnarray}
where index $n$ represents the radial harmonic index and
$ \bar{\bar{{\bf G}}}_n ({\bf r}_m,{\bf r},\omega )$ contains
contribution of the $LS{P_n}$(dipolar plasmon for $n=1$,
quadrupole plasmon for  $n=2$ , etc.). Similar to the Green
tensor, the electric field operator associated with the plasmonic
structure can be written in terms of the several multipole
components
\begin{eqnarray}\label{electric field decomposition}
{\hat {\bf E}}({\bf r}_m,\omega ) &=& \sum\limits_{n = 1}^\infty  {\hat {\bf E}_n ({\bf r}_m,\omega )}.
\end{eqnarray}
The component ${\hat {\bf E}_n ({\bf r}_m,\omega )}$
describes the electric field operator associated with the $LS{P_n}$. Point group theory tells us that the fundamental vibration of nanoribbon with the breathing-like displacement of the phonon mode has the $A_g$ symmetry (diagonal Raman tensor) just for special scattering patterns of a doubly clamped nanoribbon~\cite{R. Gillen2010,Gillen2010}. If this condition is met, the spectral decomposition of the electric field operator in the frequency domain leads to the simplified form of the interaction Hamiltonian in Eq. (2) as
\begin{eqnarray}\label{interaction Hamiltonian}
&&\hat H_{int}=  - \sqrt {\frac{\hbar }{2\omega _m}} (\hat b + \hat b^\dag)\int {\int {d\omega \,d\omega '}}\,{{\cal R}_{ii}}\\
&&\times\left\{ {\hat E_i^{(-)} ({\bf r}_m,\omega )\hat E_i^{(+)} ({\bf r}_m,\omega ')+\hat E_i^ {(+)} ({\bf r}_m,\omega )\hat E_i^{(-)} ({\bf r}_m,\omega ')} \right\}\nonumber,
\end{eqnarray}
where ${\cal R}_{ii}$  are diagonal elements of the
SGNR Raman tensor with $i=x,y,z$ (here the Einstein's sun convention is used). In the above Hamiltonian,
we have neglected two terms in Eq. (A3) since they don't satisfy the energy conservation.
Up to now, we have considered the localized plasmon polaritons in a
continuous band description. However, the dynamic of the composite
system is based on the excitation of the $LSP$
modes of the spherical nanostructure by an external pump laser and
optomechanical coupling to the vibrational mode of the mechanical
subsystem.
Therefore, based on the scheme of the mode-selective
quantization developed in Refs.~\cite{Dzsotjan2016,Castellini2018}, along with Gram-Schmidt
orthogonalization procedure, we suppose that
N discrete modes participate in the coupling process. In this manner, we
introduce mode-selective bosonic annihilation operator as
\begin{eqnarray}\label{mode selective operator }
\hat a_{n}({\bf r}_m,\omega ) = \sqrt {{{\cal R}_{ii}}/{(2\omega _m \hbar )}^{1/2}}{\alpha _n}(\omega ){\hat e_i} \hat E_{n,i}^ {(+)} ({\bf r}_m,\omega ),\nonumber\\
\end{eqnarray}
where $\hat e_i (i=1,2,3)$ are the cartesian unit vectors.
The operator $\hat a_n({\bf r}_m,\omega )$ is associated with the  $LS{P_n}$ mode at the position of the nanoribbon and can be excited by the external laser at a given position of the mechanical resonator. It is worth noting that the introduced mode-selective bosonic operators satisfy the usual bosonic commutation relation $\left[ {\hat a_{n}({\bf r}_m,\omega
),\hat a_{m}^\dag ({\bf r}_m,\omega ')} \right] = \delta
(\omega  - \omega ') \delta _{nm}$, provided that ${\alpha _n}(\omega )$  is given
by relation $\left| {\alpha _n}(\omega )\right|^2 =
1/{\left| {k_n}(r,\omega ) \right|^2}$ with
\begin{eqnarray}\label{spectral function}
\left| {{k_n}({\bf r},\omega )} \right|^2 &=& ({{\cal R}_{ii}}k_1^2/\pi {\varepsilon _0})\\
&&\times \sqrt {\hbar /2 \omega _m} \, {\hat e_i}\cdot \left[{\rm Im}{\bar{\bar{\bf G}}}_{n,s} ({\bf r}_m,{\bf r},\omega )\right]\cdot {\hat e_i}.\nonumber
\end{eqnarray}
Employing the quantization method, the interaction part of the
Hamiltonian can be obtained in terms of frequency dependent
bosonic operator $\hat a_n({\bf r}_m,\omega )$ and the spectral
coupling functions $k_{n,m}({\bf r}_m,\omega )$.
To completely perform the mode-selective quantization
scheme and obtain the final effective Hamiltonian, we need to determine the explicit form of the spectral function. Considerations related to the symmetrical and vibrational properties of the phonon modes, as
well the electromagnetic Green's tensor help us to identify the form of ${k_n}({\bf r}_m,\omega )$. For an armchair graphene ribbon in the $xy$ plane with the periodicity direction along $y$, the $A_g$ mode is Raman
active for geometry $(zXXz)$.
Here
$zz(XX)$ represent the propagation (polarizations) directions of
the incident and scattered light, respectively ~\cite{R.
Gillen2010,Gillen2010,Saito2010}. Based on the properties of the
$A_g$  symmetrized Raman tensor, we only need diagonal components
of the scattering Green's tensor $\bar{\bar{G}}_{s,
ii}^{(11)}({\bf r}_m,{\bf r}_m)$ with $i = x$. By applying
symmetry arguments, the optomechanical coupling spectra,
$K_n={{\left| {k_n(r_m,\omega )} \right|}^2}$, is obtained as following
\begin{eqnarray}\label{spectrul function}
K_n= ({{\cal R}_{ii}}\, k_1^2/\pi {\varepsilon _1})\sqrt {\hbar /2\omega _m}\, {\rm Im} \left[ {\bar{\bar G}}_{n,s}({\bf r}_m,{\bf r}_m,\omega )\right]_{ii}.\nonumber\\
\end{eqnarray}
Finding the general form of spectral function, we can go one step
further and introduce new frequency-independent plasmonic operators
$\hat a_n(r_m)$ at the position of the nanoribbon
\begin{eqnarray}\label{final mode selective operator}
\hat a_n({\bf r}_m) = \beta _n \int d\omega \, {k_n} ({\bf r}_m,\omega ) \, \hat a({\bf r}_m,\omega ).
\end{eqnarray}
The coefficient $\beta _m$ is determined through the relation ${\left| {\beta _m}
\right|^2} = 1/\int {d\omega \, {\left| {k_n(r_m,\omega )}
\right|}^2} $ so that these new-defined plasmonic operators satisfy
the bosonic commutation relation as $\left[ {\hat
a_n ({\bf r}_m),\hat a_m^\dag ({\bf r}_m)} \right] = \delta
_{nm}$. The effective Hamiltonian is obtained by integrating over the angular frequency to
establish a set of N discrete modes.

\section{Derivation of the free field Hamiltonian for plasmonic subsystem }\label{App:Green tensor}
In the following, we indicate the main steps toward deriving the quantum
Langevin equation and free field Hamiltonian of the plasmonic
subsystem. Based on the methods developed in
Refs.~\cite{Dzsotjan2016,Castellini2018}, we can decompose the plasmonic part of the
free field Hamiltonian into the predefined bright mode and a set of
continuous non-interacting orthogonal dark modes
\begin{eqnarray}\label{dark mode operator}
{\rm{\hat d}}({\bf r},\omega ) = {\rm{\hat f}}({\bf r},\omega ) - \sum\nolimits_{n = 1}^N {\left[ {\rm{\hat f}}({\bf r},\omega ),\hat a_n^\dag ({\bf r}_m,\omega ) \right] {\hat a}_n({\bf r}_m,\omega )}.\nonumber\\
\end{eqnarray}
Since dark modes have independent dynamics and will not affect the dynamics of the bright
modes, after some algebraic calculations, we arrive at the relation
\begin{eqnarray}\label{polaritonic operator based on dark and bright operators}
\int {{d^3}r \, {\rm{\hat f}}^\dag ({\bf r},\omega ){\rm{ \hat f}}({\bf r},\omega )}&=&\sum\limits_{n = 1}^N {\hat a_n^\dag ({\bf r}_m,\omega ){\hat a}_n ({\bf r}_m,\omega )}\\
&&\ +\int {{d^3}r \, {\rm{\hat d}}^\dag ({\bf r},\omega )\hat d({\bf r},\omega )}\nonumber.
\end{eqnarray}
The term related to the free Hamiltonian of dark modes can be
omitted since it plays no role in the dynamic of the system. To
determine the plasmonic part of the Hamiltonian based on the
frequency-independent bosonic operator, we first start with the
quantum Langevin equation for the new plasmonic operator, $d\hat
a_{n'}({\bf r}_m)/dt = \left[ {\hat
a_{n'}({\bf r}_m),{\hat H}_0}\right]/i\hbar $, in which the Hamiltonian ${\hat H}_0$ is defined as
\begin{eqnarray}\label{free feld Hamiltonian of plasmonic subsystem}
\hat H_0 = \int {d\omega \hbar \omega \sum\limits_{n = 1}^N {\hat a_n^\dag ({\bf r}_m,\omega ) {\hat a}_n({\bf r}_m,\omega )}}.
\end{eqnarray}
Here, we have used the definition of bosonic annihilation operator
$\hat a_n({\bf r}_m,\omega )$ in the position of the SGNR
to rewrite the Langevin equation as
\begin{eqnarray}\label{Langevin equation for modeselective operator}
{\dot{ \hat a}_{n'}}({\bf r}_m) =  - \int {d\omega \, \omega \, {k_{n'}} ({\bf r}_m,\omega ){\hat a_{n'}}({\bf r}_m,\omega )/g_{n'}({\bf r}_m)}.
\nonumber\\
\end{eqnarray}
Eq. (B4) is not yet a closed equation for $\hat
a_{n'}({\bf r}_m)$, since there is an additional integral
kernel apart from the definition of the operators $\hat
a_{n'}({\bf r}_m)$. To get
a closed-form, we subtract and add the term $( - \omega _n+
i{\Gamma _n}/2)$ to the integral kernel, and then use the relation
\begin{eqnarray}\label{spectral function}
k_n({\bf r}_m,\omega ) = \sqrt {({\Gamma _n}/2\pi )} (i{g_n}({\bf r}_m)/(\omega  - \omega _n) + i{\Gamma _n}/2).\nonumber\\
\end{eqnarray}
We arrive at the following quantum Langevin equation that includes the quantum noise and dissipation terms
\begin{eqnarray}\label{Langevin equation for modeselective operator}
{\dot{ \hat a}_{n'}}({\bf r}_m) =  - i{\omega _n}{\hat a_{n'}}({\bf r}_m) - ({\Gamma _n}/2){\hat a_{n'}}({\bf r}_m) + \hat F_{n'}({\bf r}_m).\nonumber\\
\end{eqnarray}
Here, the quantum noise is defined as $\hat F_{n'}({\bf r}_m)=
- i\int {d\omega \, \sqrt {({\Gamma _n}/2\pi )} \, \hat
a_{n'}({\bf r}_m,\omega )}$. Eq. (B6) now can be considered as the
quantum Langevin equation for the new plasmonic operator with respect to
the Hamiltonian $\hat H_p = \sum\nolimits_{n = 1}^N {\hbar {\omega _n}\hat
a_n^\dag ({\bf r}_m) \hat a_n({\bf r}_m)} $. In the end, incorporating the nanoribbon Hamiltonian and as well its interaction term in the equation of motion, we get
\begin{eqnarray}\label{Langevin equation for modeselective operator}
{\dot{ \hat a}_{n'}}({\bf r}_m) = \left[ {\hat a_{n'}({\bf r}_m),\hat H_{sys}} \right]/i\hbar  - ({\Gamma _n}/2) \hat a_{n'}({\bf r}_m)
+ \hat F_{n'}({\bf r}_m).\nonumber\\
\end{eqnarray}
%
\section{Linearization of the Quantum Langevin equations in pump-probe technique}\label{App:Green tensor}
Heisenberg equations of motion for the plasmonic annihilation operator and the mean response
of the mechanical subsystem, $\hat n = \hat b +\hat b^\dag$, are
given by
\begin{subequations}\label{Langevin equations}
\begin{eqnarray}
{\dot {\hat a}}_n ({\bf r}_m)&=& - (i{\Delta _n} + \kappa _n ) \, {\hat a}_n({\bf r}_m) + i g_{op,n} \, {\hat a}_n({\bf r}_m)\hat n \nonumber \\
&&\ +\Omega _{pu}+ {\Omega _{pr}\exp ( - i\delta t)} + {\hat F}_n({\bf r}_m),\\
\ddot {\hat n }+ \gamma \dot {\hat n}  &=&- \omega _m^2\hat n+ 2{\omega _m} \, g_{op,n}\hat a_n^\dag({\bf r}_m)  {\hat a}_n({\bf r}_m) + \hat \xi (t).\nonumber \\
\end{eqnarray}
\end{subequations}
Eq. (C1a) represents the dynamics of the $n^{th}$ plasmonic mode. The correlations associated with the
quantum vacuum fluctuations of the plasmonic cavity modes are fully characterized by the $\delta$-correlation functions as developed in Refs.~\cite{Aspelmeyer2014,Scully1997}. In Eq. (C1b), $\gamma$ and $ \hat \xi (t)$ represent the damping rate and Brownian noise of the SGNR, respectively. Here, the SGNR is affected by a Brownian stochastic force with zero mean value and the correlation function that is given in Ref.~\cite{Aspelmeyer2014}.
Since the incident probe field is much weaker than the pump field, we employ the perturbation method to investigate the optical properties of the system~\cite{Liu2018}. In the pump-probe technique,
the Heisenberg operators can be decomposed as the sum of a steady-state mean value and a small fluctuation with zero mean value, i.e., $\hat a_n (t) = \bar a_n + \delta {\hat a_n}(t)$ and $\hat n (t) = \bar n + \delta \hat n(t)$ with the steady-state mean values  $\bar a_n = \hat a_{n0}$  and $\bar n = n_0$. To obtain the steady-state solutions of the operators, we set the time derivatives of operators in the Eqs. (C1a) and (C1b) equal to zero, and yields: $a_{n0}=\Omega _{pu}/{(i(\Delta _n -
g_{op,n} n_0) + \kappa _n)}$,  and ${n_0} = 2g_{op,n} \, \omega
_0/{\omega _m}$ with $\omega _0 = {\left| a_{n0} \right|^2}$.

Inserting these definitions in Eqs. (C1a) and (C1b),
and neglecting the terms $\delta {\hat a_n}(t) \delta \hat n(t)$,
quantum Langevin equations govern the time evolution of the
fluctuation operators can be derived as
\begin{subequations}\label{Langevin equations for noise operators}
\begin{eqnarray}
\delta {{\dot {\hat a}}_n}({\bf r}_m) &=&  - (i{\Delta _n} + {\kappa _n})\delta {\hat a}_n({\bf r}_m) + i{c_0} \, g_{op,n}\delta \hat n\\
&&\ + i{n_0} \,  g_{op,n}\delta {\hat a}_n({\bf r}_m) + \Omega _{pr} \, {e^{ - i\delta t}} + {\hat F}_n({\bf r}_m) ,\nonumber
\nonumber\\
\delta \ddot {\hat n}& =&  - \gamma \, {\delta \dot {\hat n}}  - \omega _m^2\delta \hat n + 2{c_0}{\omega _m}  \, g_{op,n}\, \delta \hat a_n^\dag ({\bf r}_m)\nonumber\\
&&\ + 2c_0^ *{\omega _m}  \,  g_{op,n} \, \delta {{\hat a}_n}({\bf r}_m) + \xi (t).
\end{eqnarray}
\end{subequations}
We identify all fluctuation operators with their expectation values
and eliminate the noise terms. To solve these set of equations, we
use the following ansatz in the rotating frame
\begin{subequations}\label{ansatz }
\begin{eqnarray}
\left\langle {\delta {{\hat a}_n}({\bf r}_m)} \right\rangle  &=& {a_{n + }}\exp [ - i\delta t] + {a_{n - }}\exp [i\delta t],\\
\left\langle {\delta \hat n} \right\rangle  &=& {n_ + }\exp [ - i\delta t] + {n_ - }\exp [i\delta t].
\end{eqnarray}
\end{subequations}
Now we substitute these equations into Eqs. (C2a) and (C2b), then equate the terms with the same time dependence, and finally find the solution for $a_{n+ }$ as below
  \begin{eqnarray}\label{amplitude of probe}
 a_{n + } = \frac{\Omega _{pr}[ {w({x_n} - {y_n}) + {z_n}}]}{w\left( {x_n^2 - y_n^2} \right) + 2{y_n}{z_n}}.
\end{eqnarray}
Eq. (C4) is given in the lowest order in $\Omega _{pu}$ but to all
orders of $\Omega _{pr}$. Here, the parameter $w$, and functions $z_n
$, $y_n$ and $x_n$ are defined as $w =  - (\delta ^2 +
i\gamma \delta  + \omega _m^2)$, $z_n = 2i(\omega
_m{\omega _0) \, g_{op,n}^2}$, $y_n= i\Delta _{n} - in_0 \, g_{op,n} $ and
$x_n = \kappa _n- i\delta $. To  investigate the optical
response of the system via the output field, we employ the input-output
relation $\hat a_{n,out}(t) + \hat a_{n,in}(t) = \sqrt {2\kappa
_n} \hat a_n(t)$~\cite{Aspelmeyer2014}. Given that the
mean value of the input operator $\hat a_{n,in}(t)$ is equal to
zero, we obtain
 \begin{eqnarray}\label{output field }
\left\langle {{\hat a_{n,out}}(t)} \right\rangle & =& a_{n0,out} + a_{n + ,out} \, {e^{ - i\delta t}} + a_{n - ,out} \, {e ^{i\delta t}}\nonumber\\
 &=& \sqrt {2{\kappa _n}} \left( {a_{n0}+ a_{n + } \, {e^{ - i\delta t}} + a_{n - } \, {e ^{i\delta t}}} \right).
\end{eqnarray}
%
\section{Mie coefficients of the Green's tensor for an anisotropic sphere}\label{App:Green tensor}
The electromagnetic Green's tensor for a  radial anisotropic sphere has been extracted previously in Ref.~\cite{Qiu2007}. When the field and the source points locate in the first layer, the electromagnetic Green tensor can be separated into two parts of the vacuum and the scattering. The former describes the contribution of the direct waves from the source in the free space. The latter represents the contribution of the multiple transmission and reflection waves that occur due to interfaces. The role of anisotropy of the plasmonic structure is included in the Green tensor through a non-integer order $\upsilon= \left[ {n(n + 1)AR + 1/4}\right]^{1/2} - 1/2$ and an anisotropy ratio $AR =
\varepsilon _t/\varepsilon _r$.
By imposing the boundary conditions and employing the recursive algorithm method, all the unknown scattering Mie coefficients can be determined. For the spherical structure under study, the Mie coefficients can be obtained in terms of the transmission T-matrix elements as follows
\begin{eqnarray}\label{Mie coefficients}
B_l^{(11)} =  - T_{l,12}^{(1)}/T_{l,11}^{(1)},
\end{eqnarray}
with $l=N,M$. For $l=N$, the elements of the transformation matrix is given by~\cite{Qiu2007}
\begin{widetext}
\begin{subequations}\label{transmission matrices elements}
\begin{eqnarray}
T_{N,11}^{(1)} = \frac{{{\psi _{\upsilon}}({k_t}{R}){{\zeta '}_{n}}({k_1}{R})/{\eta _t} - {\zeta _n}({k_1}{R}){{\psi '}_{\upsilon}}({k_t}{R})/{\eta _1}}}{{{\psi _{{\upsilon }}}({k_t}{R}){{\zeta '}_{{\upsilon}}}({k_t}{R})/{\eta _t} - {\zeta _{{\upsilon}}}({k_t}{R}){{\psi '}_{{\upsilon }}}({k_t}{R})/{\eta _t}}},\\
\nonumber\\
T_{N,12}^{(1)} = \frac{{{\psi _{{\upsilon}}}({k_t}{R}){{\psi '}_{n}}({k_1}{R})/{\eta _t} - {\psi _{n}}({k_1}{R}){{\psi '}_{{\upsilon }}}({k_t}{R})/{\eta _1}}}{{{\psi _{{\upsilon}}}({k_t}{R}){{\zeta '}_{{\upsilon }}}({k_t}{R})/{\eta _t} - {\zeta _{{\upsilon }}}({k_t}{R}){{\psi '}_{{\upsilon }}}({k_t}{R})/{\eta _t}}},
\end{eqnarray}
\end{subequations}
\end{widetext}
where $k_t^2 = \omega ^2 \varepsilon _t \mu _t/c^2$, $\eta _t = \sqrt {\mu _t/\varepsilon _t}$, and $\psi_{\upsilon }(z)$ ($\zeta_{\upsilon }(z) $) represents the spherical Riccati-Bessel (Riccati-Hankel) function:
\begin{subequations}\label{Ricatti functions}
\begin{eqnarray}
\psi _{\upsilon }(z) = z \, j_{\upsilon }(z),
\\
\zeta _{\upsilon }(z) = z \, h_{\upsilon }^{(1)}(z).
\end{eqnarray}
\end{subequations}
These functions and their derivatives in the QSA are defined as
\begin{subequations}\label{Ricatti functions in QSA}
\begin{eqnarray}
&&{\psi _\upsilon }(z) \sim \frac{{{z^{\upsilon  + 1}}}}{{{2^{\upsilon  + 1}}\Gamma (\upsilon  + 3/2)}},
\\
&&{\zeta _\upsilon }(z) \sim \frac{{( - i){2^\upsilon }{{( - 1)}^{\upsilon  + 1}}\sqrt \pi  }}{{{z^\upsilon }\Gamma ( - \upsilon  + 1/2)}},
\\
&&{{\psi '}_\upsilon }(z) \sim \frac{{\sqrt \pi  (\upsilon  + 1){z^\upsilon }}}{{{2^{\upsilon  + 1}}\Gamma (\upsilon  + 3/2)}},
\\
&&{{\zeta '}_\upsilon}(z) \sim \frac{{( - i){2^\upsilon }{{( - 1)}^{\upsilon  + 1}}\sqrt \pi  ( - \upsilon )}}{{{z^\upsilon }\Gamma ( - \upsilon  + 1/2)}}.
\end{eqnarray}
\end{subequations}
Now, by substituting Eqs. (D4a) -(D4d) into Eqs. (D2a) and (D2b), the Mie coefficient $B_N^{(11)}$ is given by
\begin{eqnarray}\label{Mie coefficient}
B_N^{(11)} = \frac{i {({k_1}{R})}^{2n + 1}[(n + 1){\varepsilon _t} - (\upsilon  + 1)]}{(2n - 1) \rm{ !! }(2n + 1) \rm{ !! }\:[n \, {\varepsilon _t}+(\upsilon  + 1)]}.
\end{eqnarray}
In the first layer, i.e., the free space, the non-integer order
$\upsilon$ reduces to the integer value $n$. Moreover, the imaginary part of
the Mie coefficient $B_M^{(11)}$ vanishes, so it has no contribution to the optomechanical coupling spectra in the QSA. By defining the effective permittivity as
\begin{eqnarray}\label{permitivity}
\varepsilon _{eff}^{ani} = \varepsilon _r\left[ {{{\left[ {n(n + 1){\varepsilon _t}/{\varepsilon _r}n^2 + 1/4n^2} \right]}^{1/2}} -1/2n} \right],\nonumber\\
\end{eqnarray}
the polarizability of the anisotropic sphere takes exactly the same form as the polarizability of an effective sphere
\begin{eqnarray}\label{polarizibility}
\alpha^{ani}=  \frac{{n({\varepsilon _{eff}^{ani}} - 1)}{R}^{2n + 1} }{{n{\varepsilon _{eff}^{ani} } + (n + 1)}},
\end{eqnarray}
with effective Drude-like parameters that are introduced in section III. Now we can identify the imaginary part of
the tangential Green's tensor and the spectral function.

The resonance frequency  $\omega _n^{ani}$, which  corresponds to the $LSPn$ mode of the effective sphere in Eq. (10), can be expressed as
\begin{eqnarray}\label{plasmonic frequency}
\omega _n^{ani} = \omega _{p,eff}^{ani}\sqrt {n/[n\varepsilon _{\infty ,eff}^{ani} + (n + 1)]},
\end{eqnarray}
and the total width is $\Gamma _n^{ani} = \Gamma _p^{ani} + \Gamma _{nrad}^{ani}$. First term in $\Gamma _n^{ani}$ indicates the ohmic loss and the second term represents radiation damping rate  $\Gamma _{nrad}^{ani}$ which is given by
\begin{eqnarray}\label{plasmonic damping rate}
 \Gamma _{nrad}^{ani}=\frac{\omega _n^{ani}(2n + 1){(n + 1)({k_n}R)^{2n + 1}}}{n[n\varepsilon _{\infty, eff}^{ani} + (n + 1)](2n - 1){\rm{!! }}(2n + 1){\rm{!!}}}.\nonumber\\
\end{eqnarray}
This term is added to the ohmic loss to take into account the variation of the
electric field over the particle size~\cite{Bohren1998,Des Francs2009}.
It is obvious that for $\varepsilon _t = \varepsilon
_r$, the spherical Bessel and Hankel functions of the order $\nu$ and Drude-like parameters
reduce to, respectively, the spherical Bessel and Hankel functions of the order $n$ and the corresponding Drude parameters for the isotropic sphere. In this limit, the obtained Mie coefficients
would be the same as those for the isotropic sphere~\cite{Li1994}.


\end{document}